\documentclass[prl,aps,twocolumn,showpacs,showkeys,reprint,superscriptaddress,amsmath,amssymb]{revtex4-1}
\usepackage{graphicx}
\usepackage{color}
\usepackage{amssymb,amsmath}
\usepackage{comment}
\usepackage{ulem}
\usepackage{verbatim}

\usepackage{hyperref}
\hypersetup{
bookmarks=false,
pdftitle={Microwave photonics}
anchorcolor=blue,
colorlinks=true,
citecolor=blue,
urlcolor=blue,
linkcolor=blue}
\usepackage{marginnote}
\usepackage[author=Juanjo]{pdfcomment}
\usepackage{xcolor}

\begin{document}

\title{Circuit QED bright source for chiral entangled light based on dissipation}

\author{Fernando Quijandr\'{\i}a} 
\affiliation{Instituto de Ciencia de Materiales de Arag\'on y Departamento de F\'{\i}sica de la Materia Condensada, CSIC-Universidad de Zaragoza, Zaragoza, E-50012, Spain.}

\author{Diego Porras}
\affiliation{Departamento de F\'{\i}sica Te\'orica I, Universidad Complutense, Madrid E-28040, Spain}

\author{Juan Jos\'e Garc\'{\i}a-Ripoll}
\affiliation{Instituto de F\'{\i}sica Fundamental, IFF-CSIC, Serrano 113-bis, Madrid E-28006, Spain}

\author{David Zueco} 
\affiliation{Instituto de Ciencia de Materiales de Arag\'on y Departamento de F\'{\i}sica de la Materia Condensada, CSIC-Universidad de Zaragoza, Zaragoza, E-50012, Spain.}
\affiliation{Fundaci\'on ARAID, Paseo Mar\'{\i}a Agust\'{\i}n 36, Zaragoza 50004, Spain}

\date{\today}

\begin{abstract}

Based on a circuit QED qubit-cavity array a source of two-mode entangled microwave radiation is designed. 
Our scheme is rooted in the combination of external driving, collective phenomena and dissipation.
On top of that the reflexion symmetry is broken via external driving permitting the appearance of chiral emission.
Our findings go beyond the applications and are relevant for fundamental physics, since we show how to implement quantum lattice models exhibiting criticality driven by dissipation.
\end{abstract}

\maketitle

The controlled generation of quantum states of light is crucial for applications both in quantum information processing~\cite{Braunstein05rmp} and precision measurements~\cite{Caves80rmp}.  While Gaussian entangled states of light can be generated through non-linear optical interactions~\cite{walls.book},  the field of superconducting quantum circuits and {\it circuit-QED} provides us with more efficient tools to explore such applications \cite{Zagoskin2008}. In this respect, we recall the preparation of single mode squeezing~\cite{Castellanos-Beltran2008}, two-mode squeezing~\cite{Eichler2011b, Bergeal2012}, broadband squeezed light~\cite{Wilson2011}  single-mode nonclassical states of light~\cite{Hofheinz2009} and entanglement~\cite{Flurin2012}, as well as the extremely precise tomography both in cavities~\cite{Hofheinz2009} and in open transmission lines~\cite{Eichler2011}.

Two distinctive features of circuit-QED (cQED) are the precise positioning of quantum emitters and their local control through driving fields or, more recently, dissipation~\cite{Murch2012}. These ingredients  enable a paradigm shift in the control of light-matter interaction. First, a controlled separation between quantum emitters allows us to tailor collective effects like directionality and interference, with an accuracy that can hardly be matched by atomic systems. Second, these collective effects can now be embodied with dissipative dynamics~\cite{Marcos2012}, exploring novel dissipative quantum many-body phenomena~\cite{Diehl2008,Kraus2008,Eisert2010}, such as non-equilibrium phase transitions and dissipative criticality classes. In this work we combine both ideas, presenting a source of entangled light where the directionality and frequency of the entangled modes is tuned by means of collective phenomena, and where the source of entanglement is a dissipative many-body system that approaches criticality.

The device that we present is a quantum metamaterial that embeds a regular array of dissipative quantum emitters, merging a variety of physical effects into a single cQED device: (a) Non-perturbative sideband transitions in the qubit-cavity couplings, engineered by suitable qubit drivings. (b) An incoherent qubit re-pumping mechanism based on a combination of local reservoirs with the qubit driving. (c) Directional emission of light in the cavity array supported both by a collective interference and a the band engineering in this metamaterial. Through a detailed theoretical study of this setup we prove that our device has the following properties: (i) It acts as a source of entangled light, created by a dissipative process that arise from a combination of precisely tuned drivings with fast qubit decays. (ii) The degrees of entanglement and squeezing, as well as the frequency of the entangled modes, can be tuned by means of the periodic drivings.  (iii) It implements a chiral dissipative process with broken reflexion symmetry, induced by the phase pattern of the driving. In practice, this effect allows us to control the direction and momentum of the emitted entangled modes.  (iv) The setup can be used to study dissipative many-body collective effects, both because of its non-trivial collective steady-state, but also because the system undergoes a dissipative quantum phase transition for suitable parameters.

\begin{figure}[t]
\includegraphics[width=1.\columnwidth]{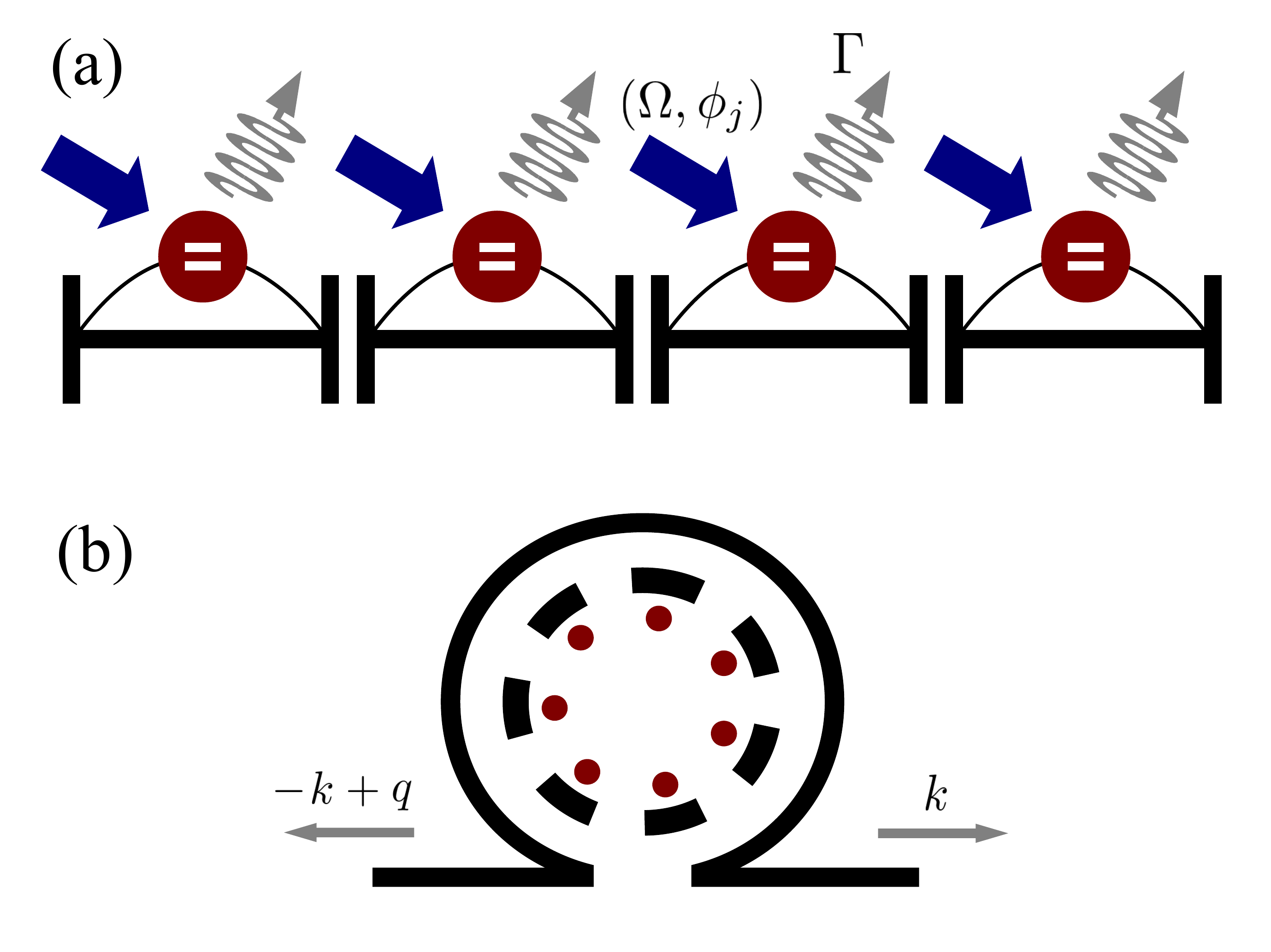}
 \caption{(a) Setup of the system studied.  Each superconducting resonator is coupled to a qubit.  The resonators are coupled each other, {\it e.g.} capacitely.  We also drawn arrows representing the classical driving and the losses.  (b) Sketch for the quantum light source. The coupled cavity can be embedded inside an open transmission line.  The entanglement generated in the array is mapped onto the line. }
 \label{fig:sketch}
 \end{figure}

\paragraph{The setup.--}
Our dissipative source of chiral quantum light is schematically drawn in Fig.~\ref{fig:sketch}a.  It consists on a 1D arrangement of coupled cavity-qubit units where the qubits gaps are periodically and uniformly driven, and where each  unit is coupled to an independent thermal bath to which they can dissipate.
We describe the coherent part of the cavity-qubit units using the Hamiltonian
\begin{equation}
\label{HJCL}
H_{{\rm CQ}} = \sum_j \left ( \omega_{\rm r} a_j^\dagger a_j + \frac{\epsilon}{2} \sigma^z_j + g \sigma^x_j X_j \right ) + J \sum_{\langle i,j\rangle}  X_j X_i 
\end{equation}
Here $\sigma_j^z$ labels a qubit state with frequency $\epsilon$, $a_j^\dagger$ represents the creation of a photon with frequency $\omega_r$, and $X_j=(a_j+a_j^\dagger)$. The Hamiltonian includes both cavity-qubit, $g$, and inter-cavity coupling, $J$. For moderate couplings and no resonant driving, the cavity-qubit and cavity-cavity interactions can be modeled in the Rotating Wave Approximation (RWA), forming the so called Jaynes-Cummings lattice (JCL), with $\sum_j g (\sigma^+_j a_j + {\rm H.c.})+\sum_{\langle ij \rangle} J (a^\dagger_i a_j + {\rm H.c.})$. 
 Analogously to the Bose-Hubbard model, it supports an insulator-superfluid quantum phase transition \cite{Hartmann2006, Greentree2006, Angelakis2007, Koch2009, Leib2010, Hummer2012}, which originates from the competition between the {\it repulsive} non-linearity induced by the two-level system and the hopping term $J$.  A promising field to implement such many-body systems is cQED, as already chips with dozens of coupled {\it identical} resonators have been reported~\cite{Underwood2012,Houck2012}.

In contrast with ordinary JCL's, for us the qubits do not act as a source of nonlinearity, but rather mediate an engineered dissipation.  Furthermore, our model strongly relies on the use and control of external qubit drivings.  All together the dynamics is given by a {\it Linblad}-type master equation~\footnote{For all purposes we can focus on the case of zero temperature use $\hbar =1$ through this paper.} that reads
\begin{align}
\label{qme0}
d_t
\varrho
=&
-i
[H_{{\rm CQ}}+ H_{\rm AC}, \varrho]
\nonumber \\
&+
\Gamma
\sum_{j}
\left ( 
\sigma^-_j \varrho \sigma^+_j -
\frac{1}{2} \{ \sigma^+_j \sigma^-_j, \varrho \} 
\right )
\, .
\end{align}
Through the paper, we neglect radiative losses, which can be straightforwardly included in our formalism and do not modify qualitatively the discussion below.
The first important ingredient in this model is a uniform two-tone driving of all qubit gaps
\begin{equation}
\label{Hdriving}
H_{\rm AC}
= \sum_{n=1,2}
\sum_j \lambda_n \cos (\Omega_n t + \phi_{n,j}) \sigma^z_j,
\end{equation}
with tunable phases, $\phi_{1,2j}$.  As we will see this driving plays a major role in the collective effects. The second important ingredient is the cavity and qubit dissipations, $\kappa$ and $\Gamma$. In this letter we work in the limit of strong qubit decay, $\Gamma\gg g$, in which the quantum emitter customizes the effective environment of the photons. 

\paragraph{Dissipation engineering.--}
Dissipation is normally regarded as the worst enemy for preserving quantum coherence in general and entanglement in particular. 
However, an appropriately engineered dissipation is a very efficient and robust way to drive a system to the desired quantum state. This powerful idea has been considered not only on single-particle models such as laser cooling~\cite{Cirac1992, Cirac1993}, but also more recently in the engineering of strongly correlated phases~\cite{Syassen2008} and distant entangled states~\cite{Krauter2011, Muschik2011}. %

The principle underlying all these examples is a tailored system-bath interaction, which allows us to engineer the dissipator $\mathcal{L}$ in the master equation governing the irreversible dynamics, $\partial_t\rho = \mathcal{L}\rho$. These dissipators drive the system of interest to its {\it stationary} state, $\mathcal{L} \varrho^*=0$.  In most relevant cases, Davies' theory assures convergence to $\varrho^*$ for any initial condition~\cite{Rivas2011}, thus avoiding the need to initialize quantum states.
All this together makes dissipation engineering a promising new paradigm for quantum information processing~\cite{Diehl2008, Verstraete2009, Kraus2008}.

Our source of entangled light is also based on engineered dissipation. In absence of driving, the setup from Fig.~\ref{fig:sketch} thermalizes via Eq.~(\ref{qme0}) to a photon vacuum. However, as soon as we modulate the qubit energy levels, the effective dissipation changes its asymptotic state, $\rho^*$,  to a product of two-mode entangled states. To make our arguments clear, we will first sketch the main ideas using a single qubit-cavity system and then introduce the photon hopping, $J$, leaving all details for the supplementary material~\cite{EPAPS}.

Following Refs.~\cite{Cirac1992, Cirac1993, Porras2012}, we prove in three steps that a bad qubit can cool a cavity to a squeezed vacuum. First, we adopt an interaction picture with respect to the qubit, resonator and driving, $\widetilde H_{\rm CQ1} = g (a \, {\rm e}^{-i \omega_{\rm r} t} + {\rm h.c.} ) ( \sigma^+ \, {\rm e}^{i f(t)} + {\rm h.c.})$, where $f(t) = \epsilon t + \sum_{\alpha = 1,2} (\lambda_\alpha/\Omega_\alpha)\sin(\Omega_\alpha t + \phi_\alpha)$  includes the periodic driving~(\ref{Hdriving}).  Next, under the assumption of weak driving, we use the Jacobi-Anger expansion~\footnote{$e^{i z \sin \theta} = \sum_{n=-\infty}^{\infty} J_n(z) e^{i n \theta}$ with $J_n(z)$ the Bessel function of first class.}, retaining terms up to order $(\lambda_\alpha /\Omega_\alpha)$. The result in the RWA is
\begin{equation}
\label{HCQ}
 \widetilde H_{\rm CQ1} = \bar g (b^\dagger \sigma^-{\rm e}^{-i \Delta t}  + {\rm h.c.})
\;
\end{equation}
for  driving frequencies $\Omega_{1,2} = \Delta + \epsilon \pm \omega_r$. Note that the qubit is now coupled to the squeezed resonator modes
\begin{equation}
\label{b}
b =   a 
+ 
\eta  {\rm e}^{i q}   a^\dagger 
\; ,
\qquad
\eta = \frac{g_2}{g_1} 
 \; \;
,
\; \;
q = \phi_1 - \phi_2 \;.
\end{equation}%
Here and in the following we consider for simplicity unnormalized squeezed modes:
$[b, b^\dagger] \neq 1$.  
The squeezing and the effective coupling strength, $\bar g = g \times g_1 {\rm e}^{-i \phi_1}$, are determined by the external driving, $g_{1}= J_{1} (2\lambda_1 / \Omega_1 ) J_{0} (2\lambda_2 / \Omega_2 )$ and $g_{2}= J_{0} (2\lambda_1 / \Omega_1 ) J_{1} (2\lambda_2 / \Omega_2 )$, through Bessel functions. The final step consists in an adiabatic elimination following the hierarchy of time scales $\epsilon \gg \Gamma \gg {\omega_{r}, g}, \Delta$. The first inequality validates Eq.~(\ref{qme0}) since it allows to treat the qubit dissipation in a weak-coupling or {\it Lindblad} type master equation.  The second inequality justifies not only the RWA in the previous steps, but it also indicates that the qubit relaxation, $\Gamma$, is faster than its interaction with the cavity, and it may be thus adiabatically eliminated. The result of these manipulations is a master equation,  $d_t\varrho =  \frac{2\bar g ^2}{\Gamma}(2b \varrho
b^{\dagger}-\{b^{\dagger} b,\varrho \})$, that cools the resonator to the vacuum of the squeezed mode, $b$.

What happens when the on-site squeezing of each resonator competes with a coupling between resonators? Since the dynamics of both processes happens on different operator basis, one would a expect a competition between these phenomena and even some phase transition. For answering this problem we now move on to the coupled Hamiltonian~(\ref{HJCL}). In the simplest case of translational invariance and periodic boundary conditions, we can introduce momentum space modes
 $a_k = N^{-1/2} \sum_j {\rm e}^{-i k j} a_j$ ($k \in 2 \pi / N \times \mathbb{Z}$), and show that the problem  is similar to the single resonator case.
More precisely, following the same definitions and approximations as before, we obtain the effective master equation in the appropiate interaction picture (see Supp. Mat. \cite{EPAPS})
\begin{align}
\label{qme}
d_t \varrho=\sum_k
-i \omega_k
[ a_k^{\dagger}a_k,\varrho] 
+ \frac{2\bar g ^2}{\Gamma}(2b_k\varrho
b_k^{\dagger}-\{b_k^{\dagger} b_k,\varrho \}).
\end{align}
The chain of qubits is now cooling the resonators to the vacuum of
two-mode-squeezed operators $b_k = a_k  + \eta  a_{-k+q}^\dagger $ with dispersion relation ($J \ll \omega_r$) $\omega_k = \Delta + 2 J \cos (k)$.
Let us remark how the external driving fully determines the properties of the asymptotic state. In particular, while the phase of the driving, $q=\phi_1-\phi_2$, selects the pairing between modes in $b_k$, we will show that the choice of frequencies, $\Delta=(\Omega_1+\Omega_2-2\epsilon)/2$, customizes the band structure and the amount of entanglement. These are the main practical results in this manuscript.

\paragraph{Entanglement in the stationary solution.---}
After the adiabatic elimination, the effective master equation~(\ref{qme}) can be written as a direct sum of quadratic dissipators acting on the Fock spaces of the operators $(b_k,b_{-k+q})$. Consequently, the asymptotic state of the master equation will be a product of Gaussian states in each of these Hilbert spaces,  $\varrho^*=\otimes\rho_{k,-k+q}^*$. Each of the final density matrices $\rho^*_{k,-k+q}$ will fully characterized by the first and second moments of the operators $R^{(k)} = \{ Q_k, P_k, Q_{-k+q}, P_{-k+q} \}$, with $Q_k = 1/\sqrt{2} (a_k^\dagger + a_k)$ and $P_k = -i/\sqrt{2} (a^\dagger_k - a_k)$. In particular, the first moments are all zero, while the second moments are conveniently grouped in the the two mode covariance matrix
\begin{equation}
  \hat\gamma_k =  \frac{1}{2}\langle R_l^{(k)} R_m^{(k)} + R_m^{(k)} R_l^{(k)} \rangle = \begin{pmatrix} \alpha & \beta_k \\ \beta_k & \alpha\end{pmatrix}.
\end{equation}
This features a diagonal matrix $\alpha = (1 + \eta^2) / 2 (1- \eta^2) {\mathbb I_{2}}$, and two nonzero off-diagonal blocks
\begin{equation}
\label{gamma}
\beta_{k} = \tfrac{\eta}{ {\mathcal E}_{k, -k +q}^2  + (1 - \eta^2)^2} 
\begin{pmatrix}
 \eta^2 - 1 &
   {\mathcal E_{k, -k +q}}  \\ 
   {\mathcal E_{k -k+q}}  &
  \eta^2 - 1
\end{pmatrix}.
\end{equation}
Note that the most relevant parameter is
\begin{align}
{\mathcal E_{k, -k+q}} &=  \Gamma (\omega_k + \omega_{-k+q}) / 4 g_1^2\\
&= \Gamma ( \Delta + 2 J \cos (q/2) \cos (k-q/2) ) / 2 g_1^2,\nonumber
\end{align}
because it determines the degree of entanglement between the pairs of modes, $(k,-k+q$). We quantify this through the Logarithmic Negativity~\cite{EPAPS}, which is
\begin{align}
  E_N = -\log_2 \left( \frac{1+\eta^2}{1-\eta^2} - \frac{2\eta}{\sqrt{
      {\mathcal E}_{k, -k +q}^2 + (1 - \eta^2)^2}} \right) \label{EN} 
\end{align}
whenever ${\mathcal E}_{k, -k +q} < (1 - \eta^2)^{3/2}/\eta$ and zero elsewhere. 

\begin{figure}[t]
\includegraphics[width=1.\columnwidth]{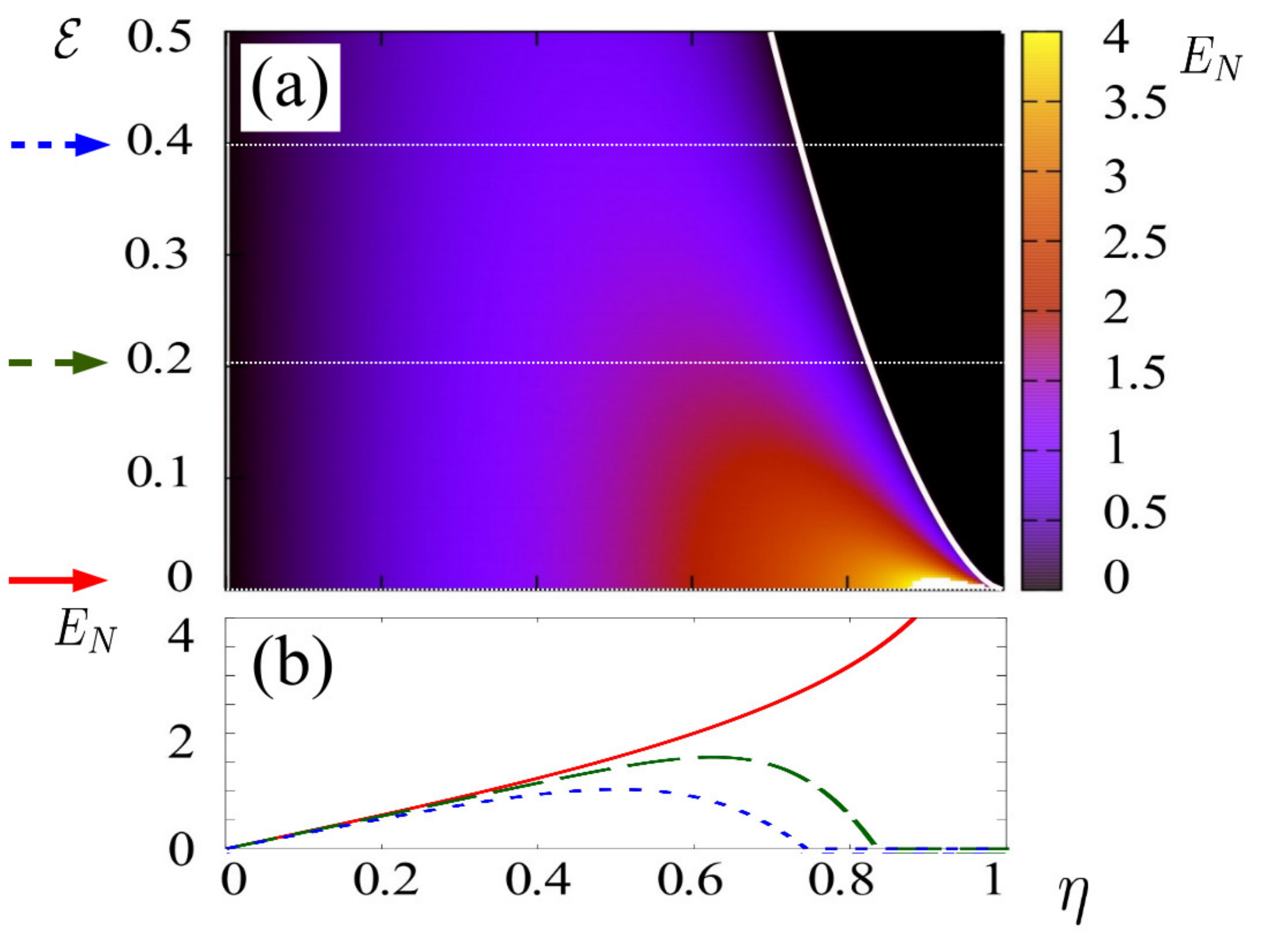}
 \caption{(a)Logarithmic Negativity $E_N$ from Eq.~(\ref{EN}).  (b) Cuts of (a) at the values of ${\mathcal E}=0,0.2$ and $0.4$ (lower), as arrow-marked in (a). For values of ${\mathcal E}$ above $0.5$ $E_N$ becomes negligible.}
 \label{fig:entanglement}
 \end{figure}

Fig.~\ref{fig:entanglement} shows a contour plot of $E_N$ in terms of both ${\mathcal  E}_{k, -k +q}$ and $\eta$, together with some sections for fixed ${\mathcal E}$.  Let us remark how entanglement significantly grows when ${\mathcal E}_{k, -k +q}$ approaches 0 and when $\eta$ approaches $1$.  A qualitative explanation follows from the master equation~(\ref{qme}): when ${\mathcal E}_{k, -k +q}$ approaches zero, it means that $\omega_k+\omega_{-k+q}=0$. Consequently, the terms that generate entanglement, $a^{\dagger}_k \rho a^\dagger_{-k+q} -\{a_k^{\dagger}a_{-k+q}^{\dagger},\varrho \} $ and  $2a_k \varrho a_{-k+q} -\{a_k a_{-k+q} ,\varrho \}$, do not oscillate and are not suppressed. Moreover, since the strength of those terms grows as $\eta^2$ and we know that $\eta\leq 1$ by definition, the optimal amount of entanglement is found when $\eta$ approaches 1.
We focus on the case $\eta \leq 1$, which corresponds to cooling to a squeezed vacuum, since the case $\eta > 1$ does not have a well defined steady-state, thus signaling an instability in the system.

We finish by an estimation of the amount of entangled mode pairs in momentum space
Fig.~\ref{fig:entanglement} shows that what really limitates the entanglement is ${\mathcal E_{k,-k+q}}$. The bigger the entanglement the closest ${\mathcal E_{k,-k+q}} $ to $0$ must be.
It turns out that  ${\mathcal E_{k,-k+q}} =0$  iff $2J/ \Delta > 1$  [Cf. below Eq. (\ref{gamma})]. 


\begin{figure}[t]
\includegraphics[width=1.\columnwidth]{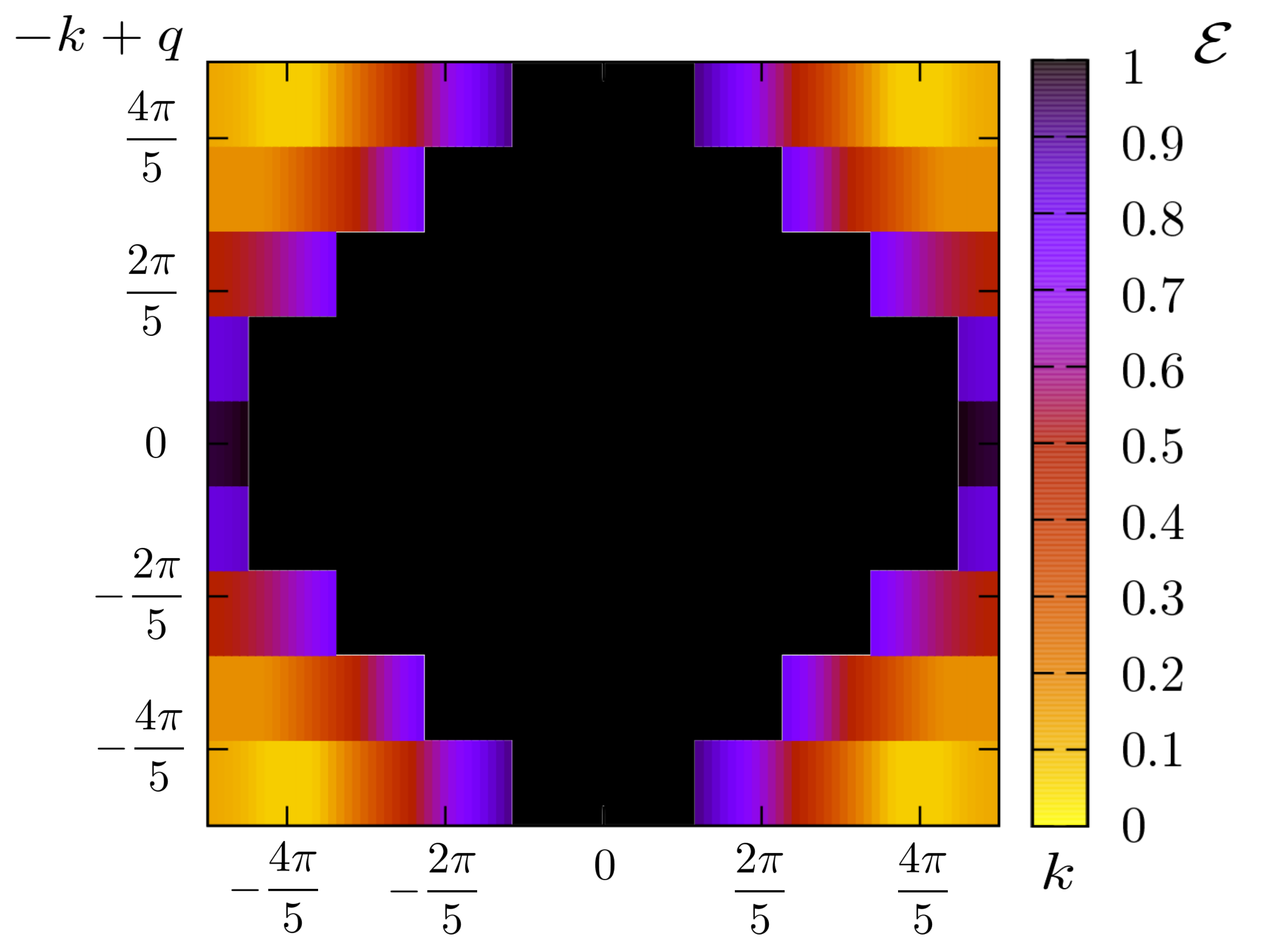}
 \caption{Contour plot of $\mathcal{E}_{k,-k+q}$ for $2J/ \Delta = 1.2$ and a lattice of $N=10$ sites.  Black zones stand for $\mathcal E \geq 1$ . 
}
 \label{fig:energy}
 \end{figure}

\paragraph{Criticality and phase transitions.---}
We have  found that entanglement diverges  in the limit $\eta \to 1$ when ${\mathcal E}_{k, -k +q} = 0 $.  In this limit it is easy to show that $\rho^*$ becomes a product of two-mode EPR state~\cite{Einstein1935}, with diverging correlation length. This is an example of a critical point which is entirely driven by dissipation~\cite{Eisert2010}.

\begin{figure}[t]
\includegraphics[width=1.\columnwidth,height=6cm]{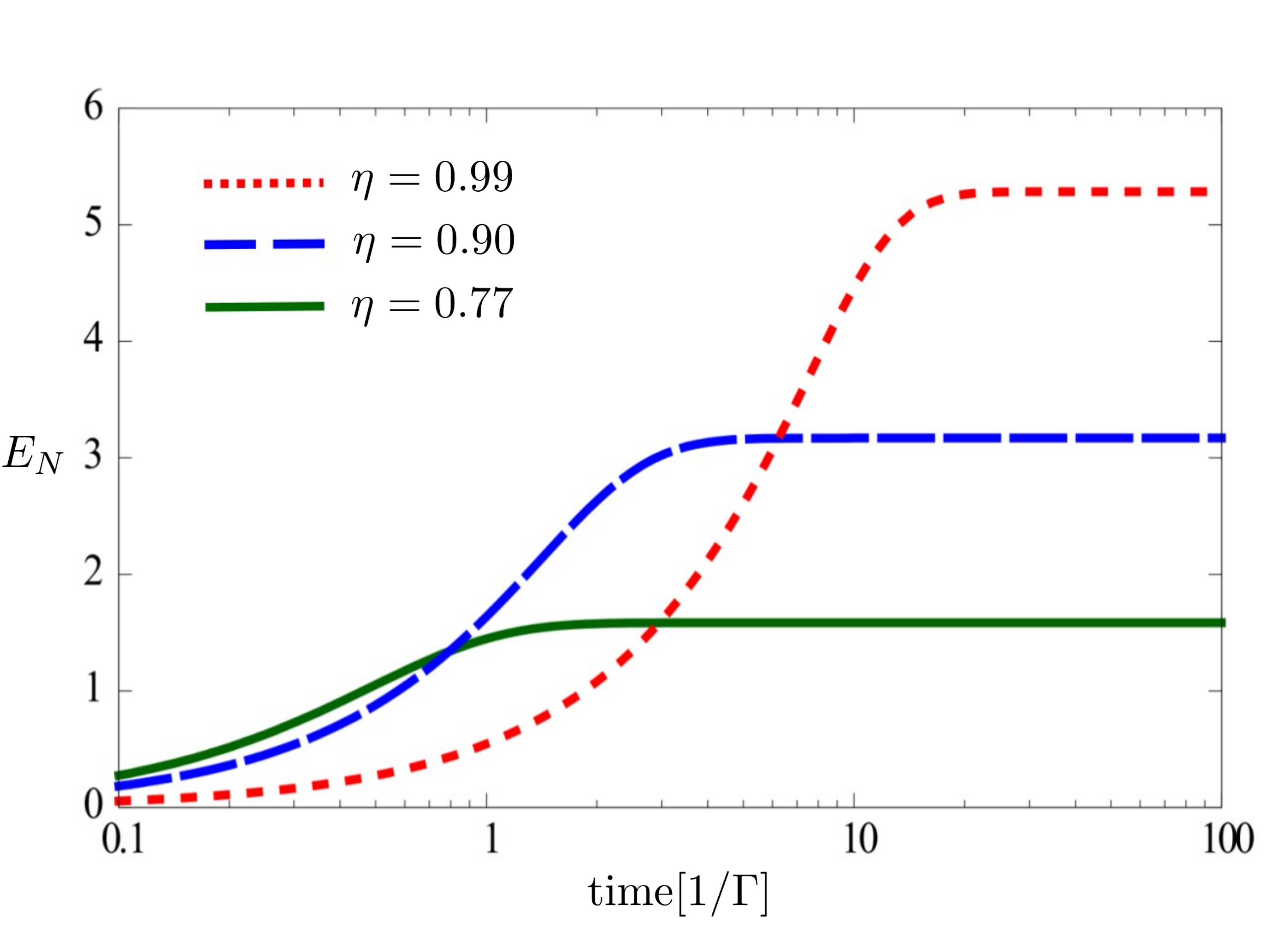}
 \caption{Time evolution of the Logarithmic Negativity $E_N$ for ${\mathcal
  E}_{k, -k +q} = 0 $ and different values $\eta$.}
 \label{fig:entanglement-time}
 \end{figure}

However, criticality in dissipative systems is characterized by the vanishing of the $\mathcal L$-eigenvalue with the largest real part (closest to zero),  ${\rm Re} (\lambda_1) \to 0$~\cite{Kessler2012}. A practical consequence is that the relaxation time becomes infinite, $T_1 \sim [{\rm Re} (\lambda_1)]^{-1} \to \infty$, close to the transition point. In our case $T_1 \sim (1 - \eta)^{-1}$.  Consequently the  rise  in the time needed to reach the stationary entangled state needs to be considered in any practical application. In other words, there is a tradeoff between the maximum entanglement achieved and the time to get it, as exemplified by the numerical simulations in Fig.~\ref{fig:entanglement-time}.

\paragraph{Quantum light emission.---}
So far, we have considered the  two mode entanglement in the coupled cavity system.
The question now is how to {\it extract} this entanglement.  
In doing so, we study the architecture depicted in Fig.~\ref{fig:sketch}b.  The resonators are surrounded by an open transmission line. The line-resonators coupling can be written as $H = i \sum_j \sqrt{\gamma/2 \pi} \int({\rm e}^{i \omega j} a_j^\dagger b(\omega) - {\rm h.c.})$, where the new parameter $\gamma$ accounts for the line-resonators coupling strength per mode. In the limit of weak coupling, the emitted light may be calculated using the input-output formalism \cite{Gardiner1985,EPAPS}. We thus obtain
\begin{equation}
b_k^{in} (t) = b_k^{out}(t) + 2\sqrt{\frac{\gamma}{N \pi^2}}{\rm e}^{-i kt}a_k,
\end{equation}
where $b_k^{in} (t) = 1/2\pi \int_{k-\delta}^{k+\delta} {\rm d} \omega {\rm e}^{-i \omega t} b_0 (\omega)$ and $\delta = \pi /N$. This is the final result of the letter.  By considering the extended coupling the resonator field modes are mapped onto the line with a resolution in frequency $\sim 1/N$ ($N$ the number of oscillators).  Therefore the system presented here works as a source of chiral quantum light.

\paragraph{Conclusions.---}
In this work we present a practical application of a recent paradigm on  non-equilibrium many body physics: {\it dissipation driven criticality}.  More precisely, we show how to use dissipation and collective phenomena as a resource for a source of chiral two-mode entangled light.

From a practical side, our proposal uses simple elements, such as qubits with a tunable gap~\cite{Paauw2009, Schwarz2012}, coupled resonators arrays~\cite{Underwood2012,Houck2012} and external drivings, which are standard in state-of-the-art circuit QED experiments. We argue that it could be thus implemented and tested using recent advances in the field of microwave state tomography~\cite{Bozyigit2010, Menzel2010, Eichler2011}. Our entangled light source also features interesting properties: it is directional, with entangled modes that can be spatially separated; it is broadband, entangling different modes and frequencies, and the only parameters that need to be tuned are the intensity and the frequency of the external drivings.

From a theoretical side, our proposal shows that circuit QED is a feasible architecture for observing dissipative many-body phenomena, with concrete examples on how to implement these ideas in the lab. We believe that the models presented here can be easily extended to include novel ingredients, such as photon-photon interactions and blockade mechanisms, which would open the door to an even large family of models, now outside the real of Gaussian states.

We acknowledge support from the Spanish DGICYT under projects FIS2009-10061 and FIS2011-25167, by the Arag\'ón (Grupo FENOL), QUITEMAD S2009-ESP-1594, and the EU project PROMISCE. DP is supported with RyC Contract No. Y200200074.

\bibliographystyle{apsrev4-1}
\bibliography{chiral}

\renewcommand{\theequation}{A\arabic{equation}}
\setcounter{equation}{0}  
\section {Appendix A: Adiabatic Elimination and the Master Equation }

In this appendix we detail how to obtain the master equation (\ref{qme}). 
We start with manipulating the coherent part. We rewrite $H_{\rm CQ}$ in Eq. (\ref{HJCL}):
\begin{equation}
H_{\rm CQ}=H_0 + H_c + H_{drive}
\end{equation}
with
\begin{align}
H_0 &= \sum_j \left( \frac{\epsilon}{2}\sigma_j^z + \omega_r a_j^{\dagger} a_j \right)\\
H_c &= \sum_j   g\sigma_j^x (a_j^{\dagger}+a_j) +  J (a_j^{\dagger} +a_j)(a_{j+1}^{\dagger}+a_{j+1})\\
H_{drive} & = \sum_j \sum_{\alpha} \lambda_{\alpha} \cos(\Omega_{\alpha}t+\phi_{\alpha j})\sigma_j^z
\end{align}
where the site-dependent driving phases are given by $ \phi_{\alpha j} = \phi_{\alpha} j $.

Expanding the resonator operators in momentum space (plane wave basis), $a_k = N^{-1/2} \sum_j {\rm e}^{-i k j} a_j$, with $k \in 2 \pi / N \times \mathbb{Z}$, we can rewrite the total Hamiltonian as
\begin{equation}
H_{\rm CQ}=H_0' + H_c' + H_{drive} 
\end{equation}
where
\begin{align}
H_0' & = \sum_j  \frac{\epsilon}{2}\sigma_j^z + \sum_k \varepsilon_k a_k^{\dagger} a_k\\
H_c' & = \sum_{k,j} g (e^{-ijk}\sigma_j^x a_k^{\dagger} + \mathrm{h.c.}) 
\end{align}
and $\varepsilon_k = \omega_r + 2 J \cos (k) $, which is valid whenever $J \ll \omega_r$.
In the interaction picture with respect to $H_0' + H_{drive}$, the interaction Hamiltonian is written as
\begin{equation}
g \sum_{k,j} \left\lbrace {\rm e}^{i j k} \left(   \sigma_j^+ {\rm e}^{2if(t)} + \sigma_j^- {\rm e}^{-2if(t)} \right) a_k {\rm e}^{-i\varepsilon_k t} + \mathrm{h.c.} \right\rbrace
\end{equation}
where the time-dependent term is given by
\begin{equation}
f(t)= \frac{\epsilon}{2} t + \sum_j \sum_{\alpha} \frac{\lambda_{\alpha}}{\Omega_\alpha} \sin(\Omega_{\alpha}t + \phi_{\alpha j})\sigma_j^z 
\end{equation}
as can easily be obtained by integration. We will select two driving frequencies $\Omega_{1}= \epsilon - (\omega_r + \Delta) $ and $\Omega_{2} = \epsilon + (\omega_r + \Delta)$. Finally, neglecting those terms rotating with $\omega_r$ we get 
\begin{equation}
{H_c}'(t) =\sum_j \left( \sigma_j^+ c_j(t) + \mathrm{h.c.} \right)
\end{equation}
with $c_j(t)$ given by
\begin{eqnarray}
c_j(t) = \sum_k (g_{1} {\rm e}^{i(k- \phi_1)j} {\rm e}^{-i(2J\cos k - \Delta) t} a_k \nonumber\\+ g_{2} e^{-i(k+ \phi_2)j} {\rm e}^{i (2J \cos k - \Delta) t} a_k^{\dagger})
\end{eqnarray}

The couplings terms $g_{1}$ and $g_{2}$ depend on the frequencies  $\Omega_{1}$ and $\Omega_{2}$ as
\begin{align}
g_{1} = g J_{1}\left(\frac{2\lambda_1}{\epsilon-\Omega_{1}}\right) J_{0}\left(\frac{2\lambda_2}{\epsilon + \Omega_{2}}\right)
\\
g_{2} = g J_{0}\left(\frac{2\lambda_1}{\epsilon-\Omega_{1}}\right) J_{1}\left(\frac{2\lambda_2}{\epsilon + \Omega_{2}}\right)
\end{align}

We will now proceed with the master equation
\begin{equation}
\dfrac{d\varrho}{dt} =  L_0 (\varrho) + L_1 (\varrho)  .
\end{equation}
Here $L_0$ describes the dissipation on the qubits induced by the bath - recall that we only take into account spontaneous emission processes, therefore
\begin{equation}
L_0 (\varrho) = \sum_j \frac{\Gamma}{2} (2 \sigma_j^- \varrho \sigma_j^+ - \sigma_j^+ \sigma_j^- \varrho - \varrho \sigma_j^+ \sigma_j^-) 
\end{equation}
while $L_1$ describes the Hamiltonian evolution of the coupling ($L_1 \varrho = -i[H_c'(t),\varrho]$).

As we want to study the dissipative dynamics induced on the resonators by the qubits with steady state $\vert 0 \rangle \langle 0 \vert$, we can eliminate adiabatically the irrelevant degrees of freedom. We start by defining the projector $P$
\begin{equation}
P\varrho = \mu \otimes \varrho_{q,ss} = \mu \otimes \varrho_{q1,ss} \otimes ... \otimes \varrho_{qi,ss} ... \otimes \varrho_{qN,ss} .
\end{equation}
Here $\mu$ describes the system of resonators and we take the steady state of the qubits $\varrho_{qi,ss} = \vert 0 \rangle_{i i}\langle 0 \vert $ as a fixed state for them. In perturbation theory up to second order in $L_1$ we get that
\begin{equation}\label{A1}
\dfrac{d\mu}{dt} = - \int_0^{\infty} d\tau \mathrm{Tr_q}[H_c'(t),e^{L_0 \tau}([H_c'(t-\tau),\mu(t)\otimes \varrho_{q,ss}])] 
\end{equation}
where the Born-Markov approximation has already been performed. Expanding the commutators in \ref{A1} and taking into account that $\sigma_j^+$ and $\sigma_j^-$ are eigenstates of the super operator $L_0$ both with eigenvalue $-\Gamma / 2$ will show that the only dependence on the integration variable $\tau$ comes through terms of the form ${\rm exp}(-(\Gamma /2 \pm i\omega_k)\tau)$, where  
$\omega_k = 2J \cos k - \Delta$. Integrating these terms we get

\begin{equation}
\int_0^{\infty} {\rm d}\tau \,{\rm e}^{-(\Gamma /2 \pm i\omega_k)\tau} = \frac{2/ \Gamma}{1+2i\omega_k /\Gamma}
\end{equation}

According to the hierarchy of energies considered in this work, 
\begin{equation}
\nonumber
\epsilon \gg \Gamma \gg {\omega_{r}, g, J, \Delta}
\; ,
\end{equation}
we will always have that $\omega_k / \Gamma \ll 1$ and therefore we can neglect the imaginary term in the denominator. Expanding the commutators in \ref{A1} also shows that the associated time evolution of the operators $a_k$ is given by $a_k \,{\rm exp}(-i(2J \cos k - \Delta)t)$. Going backwards to the Schr\"odinger picture implies cancelling out these rotating terms. We can fulfill this condition applying the following transformation
\begin{equation}
a_k \to U_0(t) a U_0^\dagger(t)
\end{equation}
with $ U_0(t) = {\rm e}^{-i \bar{H}_0 t}$,
to both sides of \ref{A1}. Here $\mathcal{O}(t)$ denotes an operator in the interaction picture. Summing over the sites $j$ we finally get the quantum master equation for our coupled set of resonators 

\begin{align}
\label{A2}
\frac{d\mu}{dt}=\sum_k
-i \omega_k
[ a_k^{\dagger}a_k,\mu] 
+ \frac{2}{\Gamma}(2b_k\varrho
b_k^{\dagger}-\{b_k^{\dagger} b_k,\mu \})
\end{align}
where $b_k = g_1 a_k + g_2 a_{-k+q}^\dagger $.

\appendix
\renewcommand{\theequation}{B\arabic{equation}}
\setcounter{equation}{0}  
\section {Appendix B: Covariance matrix and logarithmic negativity }

The QME in momentum space (\ref{qme}) yields the time evolution for the correlators:
\begin{align}
\frac{d}{dt} \langle a_k a_l \rangle = \left(-i (\omega_k + \omega_l) - \frac{4}{\Gamma}(g_1^2 - g_2^2)  \right) \langle a_k a_l \rangle \nonumber\\
- \frac{4}{\Gamma}g_1 g_2 \delta_{l,-k+q} \\
\frac{d}{dt}\langle a_k^{\dagger} a_l \rangle = \left( i(\omega_k - \omega_l)- \frac{4}{\Gamma}(g_1^2 - g_2^2) \right) \langle a_k^{\dagger} a_l \rangle \nonumber\\
+ \frac{4}{\Gamma}g_2^2 \delta_{kl} 
\end{align}
and the conjugate equation for  $\langle a_k^\dagger a_l^\dagger\rangle$.
Both the time evolution and the stationary state can be obtained. Note that if the momenta $k$ and $l$ do not satisfy the relation $l = -k + q$, the covariance matrix will be diagonal and therefore there will be no entanglement between the emitted photons.

If ${\mathcal E}_{k, -k +q} = 0 $, the time-dependent covariance matrix will take the simple form
\begin{equation}\label{B1}
\gamma_{-k,k+q}(t) =
\begin{pmatrix}
 a & 0 & -b & 0 \\ 
 0 & a & 0 & b \\
 -b & 0 & a & 0 \\
 0 & b & 0 & a
\end{pmatrix}
\end{equation}
where $a$ and $b$ are time-dependent functions defined by
\begin{eqnarray}
a(t) &=& \left( \frac{g_1^2 + g_2^2}{2(g_1^2 - g_2^2)} F(t) - 1 \right) + 1\\
b(t) &=& \frac{g_1 g_2}{g_1^2 - g_2^2} F(t)
\end{eqnarray}
with $F(t) = 1-{\rm exp}\left( -\frac{4}{\Gamma}(g_1^2-g_2^2)t \right)$. The logarithmic negativity is defined as
\begin{equation}\label{B2}
E_N = -\frac{1}{2}\sum_{i=1}^4 \log_2 [\min(1,2\vert l_i \vert)]
\end{equation}
where $l_i$ are the symplectic eigenvalues of the covariance matrix. The symplectic spectrum corresponds to the eigenvalues of the matrix $i \Omega \gamma _{k, -k +q}^{T_A}$, where $\Omega$ is the symplectic matrix 
\begin{equation}
\Omega=
\begin{pmatrix}
 0 & 1 & 0 & 0 \\ 
 -1 & 0 & 0 & 0 \\
 0 & 0 & 0 & 1 \\
 0 & 0 & -1 & 0
\end{pmatrix}
\end{equation}
and the superindex $T_A$ denotes the partial transpose. This can be obtained by the transformation $\gamma _{k, -k +q}^{T_A} = \Lambda^T \gamma _{k, -k +q} \Lambda$ with $\Lambda$ defined as
\begin{equation}
\Lambda=
\begin{pmatrix}
 1 & 0 & 0 & 0 \\ 
 0 & 1 & 0 & 0 \\
 0 & 0 & 1 & 0 \\
 0 & 0 & 0 & -1
\end{pmatrix}
\end{equation}
From definition \ref{B2} and matrix \ref{B1} we finally get
\begin{equation}
E_N(t) = -\log_2 \left[ \left(\frac{1-\eta}{1+\eta}\right)F(t)+1-F(t) \right]
\end{equation} 
($\eta = g_2 / g_1$). 

Following a similar approach for the general case ${\mathcal E}_{k, -k +q} \neq 0 $ we get the steady state relation \ref{EN}.

\appendix
\renewcommand{\theequation}{C\arabic{equation}}
\setcounter{equation}{0}  
\section {Appendix C: Input-output formalism }

We consider here the interaction of the coupled qubit-cavity (\ref{HJCL}) and an infinite transmission line (TL). The total Hamiltonian for this system is:
\begin{equation}\label{C1}
H = H_{\rm CQ} + H_{\rm TL} + H_{\rm int}
\end{equation}
here $H_{\rm CQ}$ stands for (\ref{HJCL})  and $H_{\rm TL}$ describes the transmission line
\begin{equation}\label{C2}
H_{TL} = \int_{-\infty}^{\infty} d\omega \omega b^{\dagger}(\omega) b(\omega)
\end{equation}
(where the $b$ operators satisfy the commutation relation $[b(\omega'),b^{\dagger}(\omega)] = \delta(\omega' - \omega)$)
and $H_{\rm int}$ is the interaction Hamiltonian - expanding the $a_j$ operators in momentum space (plane wave basis) it is given by
\begin{equation}\label{C3}
H_{\rm int} = \frac{i }{\sqrt{N}} \sum_{jk} g(\omega) \int d\omega  \left( e^{i (\omega -k) j} a_k^{\dagger} b(\omega) - {\rm h.c.} \right)
\end{equation}
Approximating the sum
\begin{equation}
\nonumber
\sum_j e^{i (\omega -k) j}
\; ,
\end{equation}
to a rectangle of height $N$  centered at $k$ with a width of $2\delta = 2\pi /N$ (being zero elsewhere), we can rewrite \ref{C3} as
\begin{equation}\label{C4}
H_{int} \sim \frac{i}{\sqrt{N}} \sum_k \lbrace \int_{k-\delta}^{k+\delta} d\omega N g(\omega) a_k^{\dagger} b(\omega) - {\rm h.c.} \rbrace
\end{equation}

From \ref{C1}, \ref{C2} and \ref{C4} the Heisenberg equation of motion for the operator $b(\omega)$ reads,
\begin{equation}\label{C5}
\dot{b}(\omega) = -i \omega b(\omega) - \sqrt{\gamma N/2\pi} a_k
\; .
\end{equation}
Here, the first Markov approximation $g(\omega) = g = \sqrt{\gamma/2\pi}$ was assumed. Integrating \ref{C5} from $t_0$ to $t$ ($t_0<t$) gives
\begin{equation}\label{C6}
b(\omega) = e^{-i\omega(t-t_0)}b_0(\omega) - \sqrt{\gamma N/2\pi} \int_{t_0}^t dt' e^{-i\omega(t-t')} a_k(t')
\end{equation}
Integrating now \ref{C6} over the frequency interval $[k-\delta,k+\delta]$ we obtain
\begin{eqnarray}\label{C7}
\int d\omega b(\omega) = \sqrt{2\pi}b_{in}(t) - \sqrt{\frac{\gamma N}{2\pi}}  \int_{t_0}^t dt' \frac{2}{(t-t')} \nonumber\\ \times e^{-i k(t-t')} \sin(\delta(t-t')) a_k(t')
\end{eqnarray}
with $b_{in}$ defined as (following~\cite{Gardiner1985})
\begin{equation}\label{C8}
b_{in}(t) = \frac{1}{2\pi} \int_{k-\delta}^{k+\delta} d\omega  e^{-i\omega(t-t_0)} b_0(\omega)
\end{equation}

In a similar fashion we can integrate \ref{C5} from $t$ to $t_1$ ($t<t_1$) and define a corresponding \textit{out} operator. We will find that the \textit{in} and \textit{out} operators are related by
\begin{eqnarray}\label{C9}
b_{in}(t) = b_{out}(t) + \frac{\sqrt{\gamma N}}{\pi} \int_{t_0}^t dt' \frac{2}{(t-t')} \nonumber\\ \times e^{-i k(t-t')} \sin(\delta(t-t')) a_k(t')
\end{eqnarray}

In the continuum limit ($\delta \rightarrow 0$), taking $t_0 \rightarrow -\infty$ and for sufficiently long values of $t$, \ref{C9} yields

\begin{equation}\label{C10}
b_{in}(t) = b_{out}(t) + \sqrt{\frac{\gamma}{N \pi^2}} {\rm e}^{-ikt} a_k
\end{equation} 

with $a_k$ defined by

\begin{equation}\label{C11}
a_k = \frac{1}{2\pi}\int_{-\infty}^{\infty} dt' {\rm e}^{ikt'} a_k(t')
\end{equation}

\end{document}